\documentclass[12pt]{iopart}
\usepackage{iopams}
\usepackage{graphicx}
\usepackage{subfig}
\usepackage{comment}
\expandafter\let\csname equation*\endcsname\relax
\expandafter\let\csname endequation*\endcsname\relax
\usepackage{amsmath}

\begin{document}

\title{Exploiting higher-order resonant modes for axion haloscopes}

\author[kaist]{Jinsu Kim$^{1,2}$, SungWoo Youn$^{2, *}$, Junu Jeong$^{1,2}$, Woohyun Chung$^2$, Ohjoon Kwon$^2$, Yannis K. Semertzidis$^{1,2}$}
\address{$^1$Center for Axion and Precision Physics Research, Institute for Basic Science, Daejeon 34047, Republic of Korea}
\address{$^2$Department of Physics, Korea Advanced Institute of Science and Technology, Daejeon 34141, Republic of Korea}
\ead{*swyoun@ibs.re.kr}
\vspace{10pt}
\begin{indented}
\item[]\today
\date{\today}
\end{indented}

\begin{abstract}
The haloscope is one of the most sensitive approaches to the QCD axion physics within the region where the axion is considered to be a dark matter candidate.
Current experimental sensitivities, which rely on the lowest fundamental TM$_{010}$ mode of a cylindrical cavity, are limited to relatively low mass regions.
Exploiting higher-order resonant modes would be beneficial because it will enable us to extend the search range with no volume loss and higher quality factors.
This approach has been discarded mainly because of the significant degradation of form factor, and difficulty with frequency tuning.
Here we introduce a new tuning mechanism concept which both enhances the form factor and yields reasonable frequency tunability.
A proof of concept demonstration proved that this design is feasible for high mass axion search experiments.
\end{abstract}

\vspace{2pc}

\section{Introduction}
The axion is a consequence of the PQ mechanism proposed to solve the strong-CP problem in particle physics~\cite{paper:PQ}. 
If the mass falls within a certain range, it has cosmological implications as cold dark matter~\cite{paper:CDM}. 
A methodological approach to detect the axion signal employs cavity haloscopes, where, under a strong magnetic field, axions are converted to microwave photons resonating with a cavity mode~\cite{paper:Sikivie}.
The axion-to-photon conversion power is given by
\begin{equation*}
P_{a\rightarrow\gamma\gamma} = g_{a\gamma\gamma}^2 \frac{\rho_a} {m_a}B_0^2VC\,{\rm min}(Q_L,Q_a),
\end{equation*}
where $g_{a\gamma\gamma}$ is the axion-to-photon coupling, $\rho_a$ is the local halo density, $m_a$ is the axion mass, $B_0$ is the external magnetic field, $V$ is the cavity volume, $C$ is the form factor, and $Q_L$ and $Q_a$ are the cavity (loaded) and axion quality factors.
As the axion mass is a priori unknown, all possible mass ranges need to be scanned.
From the experimental point of view, the figure of merit is the scan rate, which is written as
\begin{equation*}
\frac{df}{dt} = \left(\frac{1}{\rm{SNR}}\right)^2\left(\frac{P_{a\rightarrow\gamma\gamma}}{k_BT_{\rm{sys}}}\right)^2  \frac{Q_a}{Q_L},
\label{eq:scan_rate}
\end{equation*}
where SNR is the signal-to-noise ratio, $k_B$ is the Boltzmann constant, and $T_{\rm{sys}}$ is the noise temperature of the system.

The form factor has dependence on the cavity geometry and resonant mode:
\begin{equation}
C \equiv \frac{\left| \int_V \vec{E}_c\cdot \vec{B}_0 d^3x\right|^2}{\int_V \epsilon(x) |\vec{E}_c|^2 d^3x \int_V |\vec{B}_0|^2 d^3x},
\label{eq:form_factor}
\end{equation}
where $\vec{E}_c$ is the electric field of the cavity mode under consideration, and $\epsilon(x)$ is the dielectric constant inside the cavity volume.
For cylindrical cavities, the TM$_{010}$ mode is conventionally considered since it yields the largest form factor in a homogeneous static magnetic field~\cite{paper:review}.

To date, the axion haloscope is the most promising approach with sensitivity to the QCD axion models~\cite{paper:KSVZ,paper:DFSZ}in the mass range between $10^0$ and $10^2\,\mu$eV, where the axion is considered to be a candidate for cold dark matter.
However, current experimental sensitivities are limited to relatively low mass regions since cavity-based experiments typically employ a single resonant cavity for a large detection volume and adopt the lowest resonant mode for the maximum form factor~\cite{paper:ADMX,paper:HAYSTAC, paper:ORGAN,paper:CAPP}.
Some cavity designs have been proposed to efficiently explore higher mass regions with minimal volume loss while relying on the same resonant mode~\cite{paper:multiple-cavity,paper:multiple-cell}.

\begin{table}[b]
\caption{Parameters of a cylindrical cavity for different resonant modes: resonant frequency ($f$), volume ($V$), quality factor ($Q$), and form factor ($C$).
$f$, $V$, and $Q$ are values relative to those of the TM$_{010}$ mode, while $C$ is the absolute value.}
\label{tab:parameter}
\begin{indented}
\item[]\begin{tabular}{c|cccc}
\br
~~~Mode~~~ & $f$ & $~~~V~~~$ & $Q$ & $~~~C~~~$ \\
\mr
TM$_{010}$ & $f_{010}$ & $V_{010}$ & $Q_{010}$ & 0.69 \\ 
TM$_{020}$ & $2.3\times f_{010}$ & $V_{010}$ & $1.5\times Q_{010}$ & 0.13 \\ 
TM$_{030}$ & $3.6\times f_{010}$ & $V_{010}$ & $1.9\times Q_{010}$ & 0.05 \\ 
\br
\end{tabular}
\end{indented}
\end{table}

\begin{figure}[t]
\centering
\includegraphics[width=0.6\textwidth]{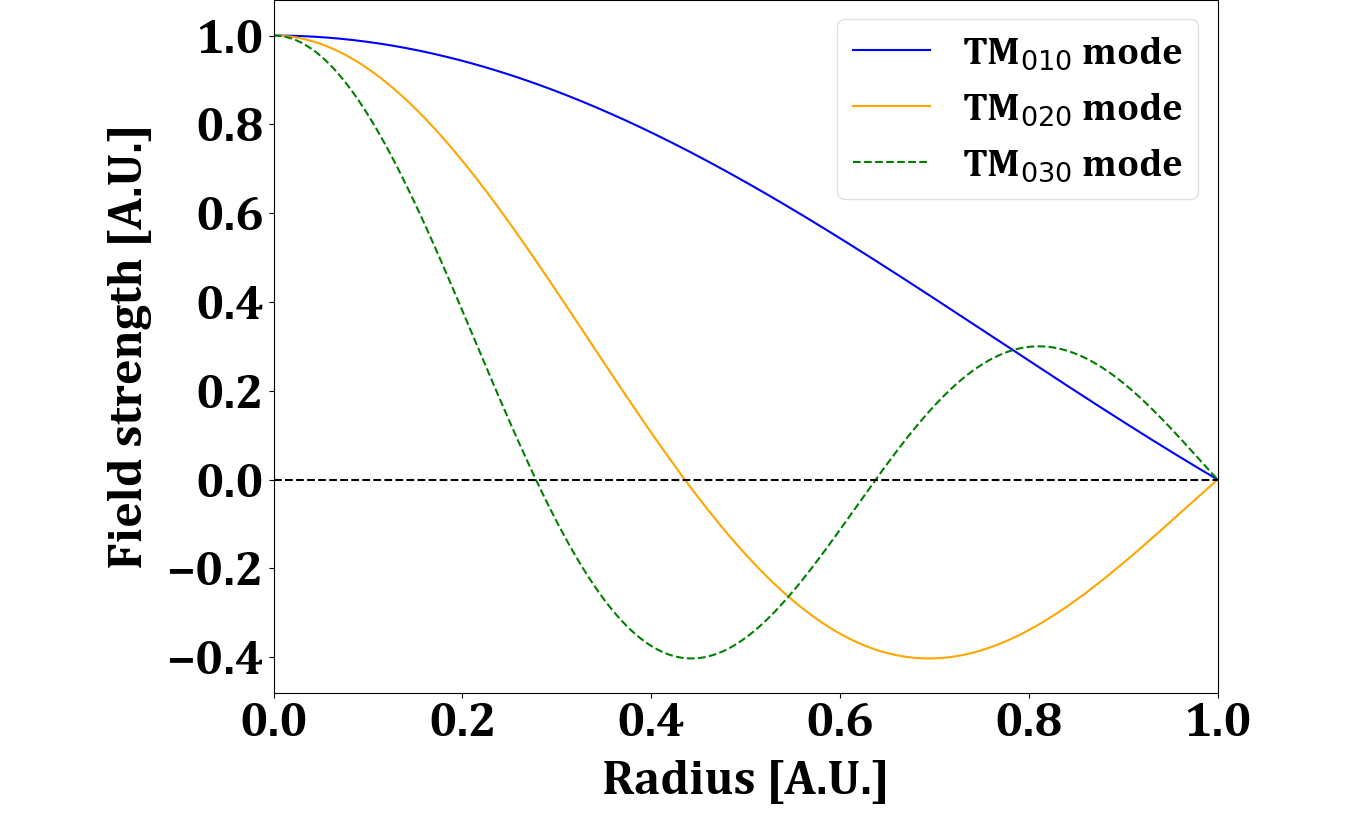}
\caption{Electric field profiles (in the $rz$ plane) of the TM modes, listed in Table~\ref{tab:parameter}, of a cylindrical cavity.
The maximum field strength and cavity radius are normalized to unity.}
\label{fig:eprofiles}
\end{figure}

\begin{figure}[b]
\centering
\subfloat[]{\includegraphics[width=0.55\textwidth]{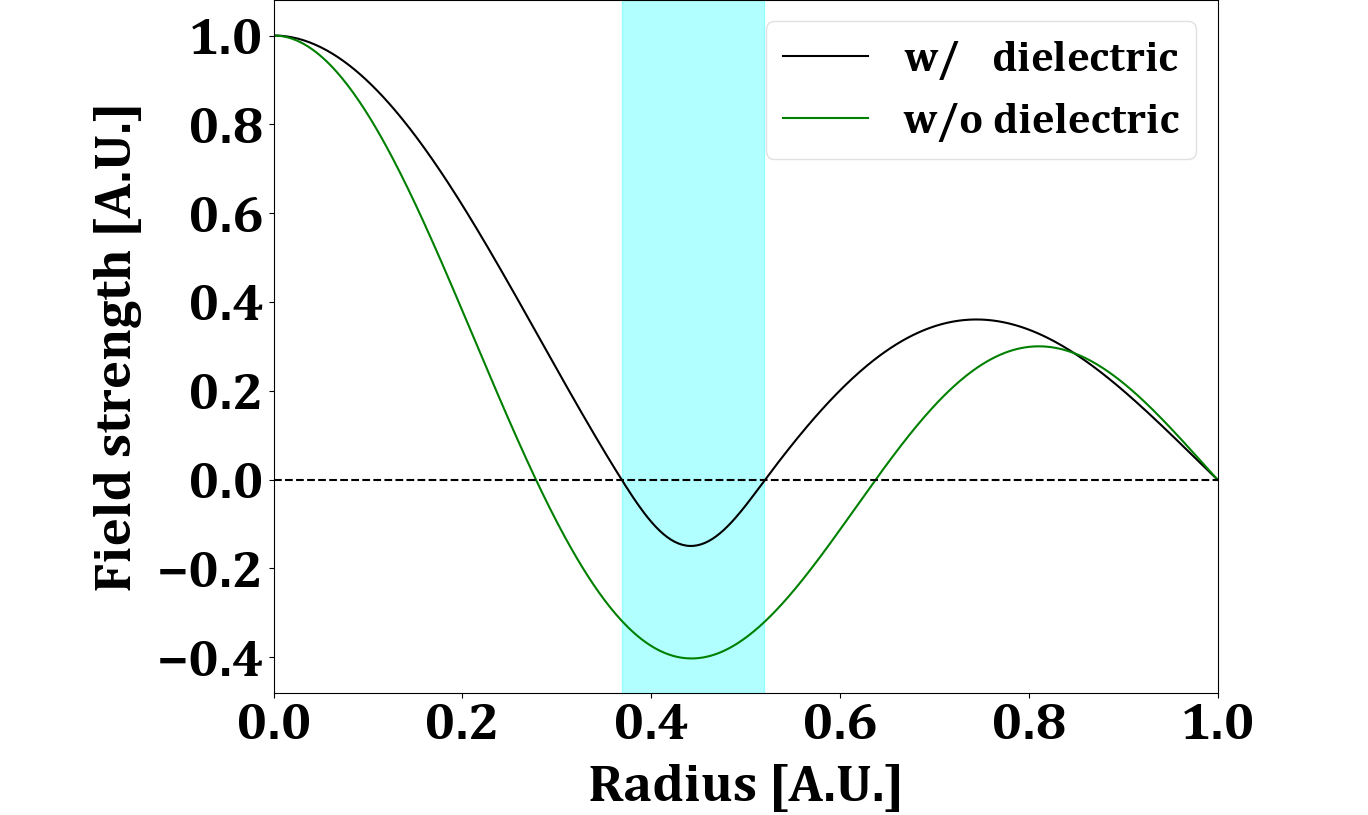}}
\subfloat[]{\includegraphics[width=0.33\textwidth]{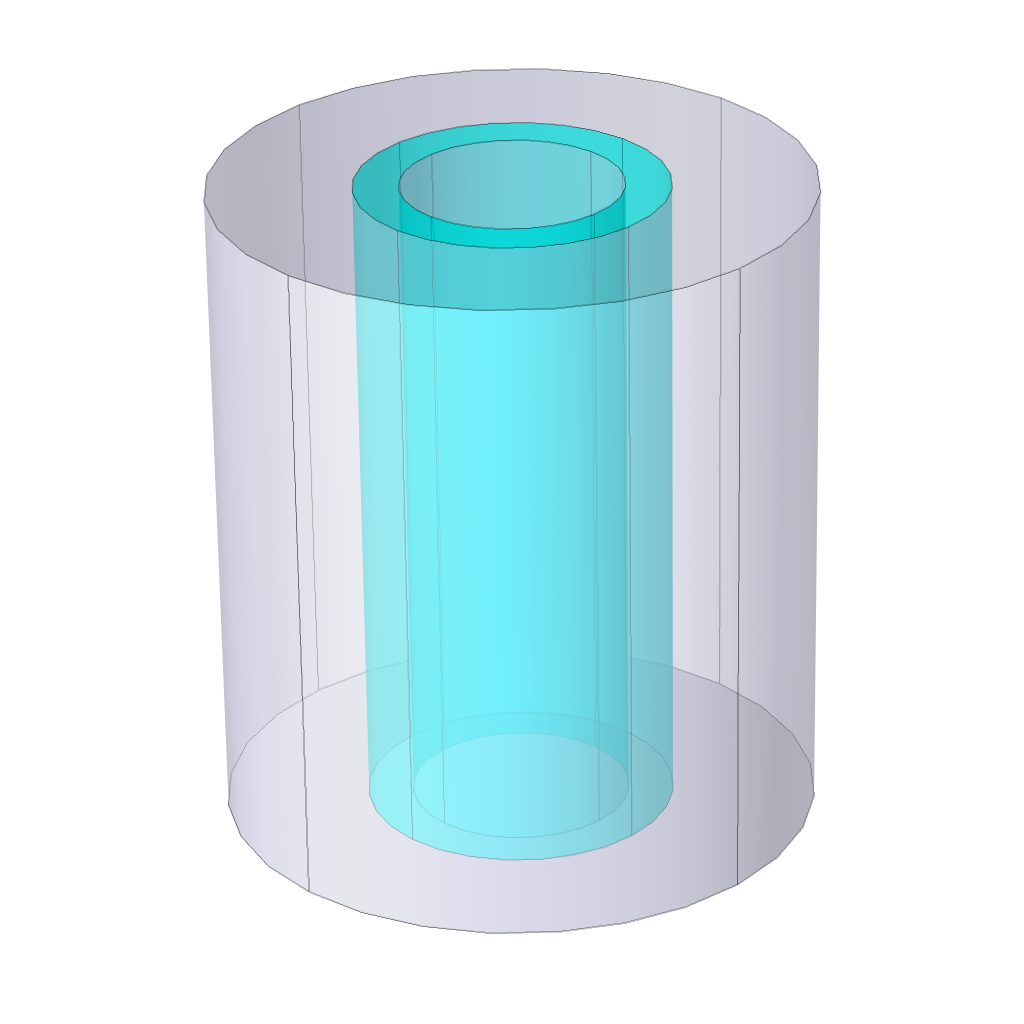}}
\caption{(a) 
Effect of a dielectric medium (in cyan) on the electric field of the TM$_{030}$ mode.
The black solid and green dashed lines are the electric field profiles with and without the medium. 
(b) A dielectric hollow structure for a cylindrical cavity to implement the effect.
}
\label{fig:dielectric_effect}
\end{figure}

As an alternative method to extend the search range towards higher mass regions, it could be beneficial to exploit the higher-order resonant modes in a cylindrical cavity.
This would enable us to access higher frequency regions without volume loss and even with higher quality factors, as summarized in Table~\ref{tab:parameter}.
However, as shown in Fig.~\ref{fig:eprofiles}, high-degree variations in the cavity EM field give rise to out-of-phase field components, which, under an external magnetic field, results in negative contributions to the form factor in Eq.~\ref{eq:form_factor}.
The negative effect becomes larger with the increasing order of the resonant mode, as can be seen in Table~\ref{tab:parameter}.
This significantly reduces the experimental sensitivity, and consequently, the higher modes have not been considered for axion search experiments.
However, a (periodic) structure of dielectric material can suppress the out-of-phase electric field component(s), which would enhance the form factors substantially, making reasonable sensitivities achievable.
For cylindrical cavities, a (periodic) layer(s) of dielectric hollow(s) can be considered for this purpose.
Figure~\ref{fig:dielectric_effect} provides an example of such a dielectric effect and intuitive design for the TM$_{030}$ mode of a cylindrical cavity.
In this article, we examine several tuning mechanisms utilizing higher-order resonant modes, in particular the TM$_{030}$ mode, to find a suitable approach for high mass axion searches.

The optimal size and position of the dielectric material can be determined analytically by solving the EM field solution, and requiring the experimental sensitivity to be maximized.
The tunability of resonant frequency, in the meantime, also matters in order to achieve a large coverage of axion masses for a given cavity dimension.
In Sec.~\ref{sec:analytic_solution}, the general solutions for TM modes of a cylindrical geometry with a dielectric hollow are obtained and the TM$_{030}$ mode is considered for further study. 
In Sec.~\ref{sec:mechanisms}, we examine several frequency tuning schemes by breaking the field symmetry along each direction of the cylindrical coordinate system, in the longitudinal, radial and azimuthal directions.
Finally, a simple and realistic tuning mechanism is chosen to demonstrate its feasible application to axion haloscopes in Sec.~\ref{sec:demonstration}.

\section{Analytic solutions for a cylindrical cavity with a dielectric hollow}
\label{sec:analytic_solution}
\subsection{EM solutions}
\label{sec:em_solution}
The general electromagnetic field solutions for the TM modes of a cylindrical cavity, where the field symmetry is conserved along the longitudinal and azimuthal directions, are given in the cylindrical coordinate system $(r, \phi, z)$ by
\begin{equation}
\begin{aligned}
\vec{E} &= (A J_0 (\sqrt{\epsilon}kr) + B Y_0 (\sqrt{\epsilon}kr))\hat{z}, \\
\vec{B} &={\sqrt{\epsilon} \over c}[ A  J_1 (\sqrt{\epsilon}kr) + B  Y_1 (\sqrt{\epsilon}kr)] \hat{\phi},
\label{eq:gen_sol}
\end{aligned}
\end{equation}
where $J_{\alpha}$ and $Y_{\alpha}$ are the Bessel's functions of the first and second kinds for the non-negative integer $\alpha$, $\epsilon$ is the dielectric constant, and $k$ is the wave number in vacuum. 
$A$ and $B$ are coefficients to be determined by boundary conditions. 
With a cylindrical dielectric hollow in the cavity as seen in Fig.~\ref{fig:dielectric_effect}(b), there are three distinct regions - inner, medium, and outer regions, which are denoted by the subscripts $i$, $m$, and $o$, respectively, with the inner and outer regions assumed to be in vacuum, i.e., $\epsilon=1$.

For the inner region, $B_i=0$ is required as $Y_0$ diverges at the origin.
The electric field amplitude is normalized to unity, i.e., $A_i=1$, such that the coefficients in the other regions are scaled accordingly.
In the medium region filled with a dielectric material with $\epsilon>1$, the continuity of the fields at the inner and outer surfaces is imposed by the boundary conditions 
\begin{eqnarray*}
J_0 (kr_i) &= A_m J_0 (\sqrt{\epsilon}kr_i) + B_m Y_0 (\sqrt{\epsilon}kr_i ), \\
J_1 (kr_i) &= \sqrt{\epsilon} A_m J_1 (\sqrt{\epsilon}kr_i) + \sqrt{\epsilon} B_m Y_1 (\sqrt{\epsilon}kr_i),
\end{eqnarray*}
and
\begin{equation}
\begin{aligned}
A_m J_0 (\sqrt{\epsilon}kr_o) + B_m Y_0(\sqrt{\epsilon}kr_o) = A_o J_0 (kr_o) + B_o Y_0 (kr_o), \\
\sqrt{\epsilon} [A_m J_1 (\sqrt{\epsilon}kr_o) + B_m Y_1 (\sqrt{\epsilon}kr_o) ]= A_o J_1 (kr_o) + B_o Y_1 (kr_o),
\label{eq:bound_cond}
\end{aligned}
\end{equation}
where $r_i$ and $r_o$ are the inner and outer radii of the dielectric hollow.
It is straightforward to solve Eq.~\ref{eq:bound_cond} to yield
\begin{eqnarray*}
A_m &= {J_0 (kr_i) Y_1(\sqrt{\epsilon}kr_i) - J_1 (kr_i) Y_0 (\sqrt{\epsilon}kr_i) /\sqrt{\epsilon} \over J_0 (\sqrt{\epsilon}kr_i) Y_1(\sqrt{\epsilon}kr_i) - J_1 (\sqrt{\epsilon}kr_i) Y_0 (\sqrt{\epsilon}kr_i)}\\
        &=-\frac{\pi}{2} k r_{i} \left[\sqrt{\epsilon}J_{0}(k r_{i})Y_{1}(\sqrt{\epsilon}k r_{i})-J_{1}(k r_{i})Y_{0}(\sqrt{\epsilon}k r_{i}) \right], \\
B_m &= - {J_0 (kr_i) J_1(\sqrt{\epsilon}kr_i) -  J_0 (\sqrt{\epsilon}kr_i)J_1 (kr_i) /\sqrt{\epsilon} \over J_0 (\sqrt{\epsilon}kr_i) Y_1(\sqrt{\epsilon}kr_i) - J_1 (\sqrt{\epsilon}kr_i) Y_0 (\sqrt{\epsilon}kr_i)} \\
        &=\frac{\pi}{2} k r_{i} \left[\sqrt{\epsilon}J_{0}(k r_{i})J_{1}(\sqrt{\epsilon}k r_{i})-J_{1}(k r_{i})J_{0}(\sqrt{\epsilon}k r_{i}) \right],
\end{eqnarray*}
and 
\begin{equation*}
\begin{split}
A_o &= {J_0 (\sqrt{\epsilon} k r_o) Y_1 (kr_o) - J_1 (\sqrt{\epsilon} k r_o) Y_0 (kr_o) \sqrt{\epsilon} \over J_0 (kr_o) Y_1 (kr_o) - J_1 (kr_o)Y_0 (kr_o)} A_m \\ 
       &\quad +  {Y_0 (\sqrt{\epsilon} k r_o) Y_1 (kr_o) - Y_1 (\sqrt{\epsilon} k r_o) Y_0 (kr_o) \sqrt{\epsilon} \over J_0 (kr_o) Y_1 (kr_o) -  J_1 (kr_o)Y_0 (kr_o)} B_m\\
       &= -\frac{\pi}{2}kr_o \left\{ \left[ J_0 (\sqrt{\epsilon} k r_o) Y_1 (kr_o) - J_1 (\sqrt{\epsilon} k r_o) Y_0 (kr_o) \sqrt{\epsilon} \right] A_m \right.\\
       &\quad \left. + \left[ Y_0 (\sqrt{\epsilon} k r_o) Y_1 (kr_o) - Y_1 (\sqrt{\epsilon} k r_o) Y_0 (kr_o) \sqrt{\epsilon} \right] B_m \right\},\\
B_o &= - {J_0 (\sqrt{\epsilon} k r_o) J_1 (kr_o) - J_1 (\sqrt{\epsilon} k r_o) J_0 (kr_o) \sqrt{\epsilon} \over J_0 (kr_o) Y_1 (kr_o) - J_1 (kr_o)Y_0 (kr_o)} A_m \\
       &\quad  -  {J_1 (kr_o)Y_0 (\sqrt{\epsilon} k r_o)  - J_0 (kr_o) Y_1 (\sqrt{\epsilon} k r_o) \sqrt{\epsilon} \over J_0 (kr_o) Y_1 (kr_o) - J_1 (kr_o)Y_0 (kr_o)} B_m\\
       &= \frac{\pi}{2}kr_o \left\{ \left[ J_0 (\sqrt{\epsilon} k r_o) J_1 (kr_o) - J_1 (\sqrt{\epsilon} k r_o) J_0 (kr_o) \sqrt{\epsilon} \right] A_m \right.\\
       &\quad \left. + \left[ J_1 (kr_o)Y_0 (\sqrt{\epsilon} k r_o)  - J_0 (kr_o) Y_1 (\sqrt{\epsilon} k r_o) \sqrt{\epsilon} \right] B_m \right\}.
\end{split}
\end{equation*}

The boundary condition for a perfect electric conductor requires a vanishing tangential component of the $E$-field, i.e. $E_z(r=R)=0$, for which only TM$_{0n0}$ modes are allowed.
This yields from Eq.~\ref{eq:gen_sol}
\begin{equation}
A_o J_0 (k_nR) + B_o Y_0(k_nR) = 0,
\end{equation}
from which the resonant frequency is extracted as $f_{0n0}=\omega_n /2\pi = k_nc/ 2\pi$.
In this article, we particularly consider $n=3$, i.e., TM$_{030}$ resonant mode, for further study.

\subsection{Scan rate factor}
\label{subsec:scan_rate_factor}
The optimal thickness of the dielectric hollow, $r_o-r_i$, is determined by maximizing a physical quantity, $V^2C^2Q$.
This quantity is relevant to the experimental sensitivity in Eq.~\ref{eq:scan_rate} and called the scan rate factor in this manuscript.
The individual parameters can be expressed in terms of the electric field $E$ in the following manner.

For a cylindrical cavity with radius $R$ and length $L$, the volume is $V=\pi R^2 L$.
The form factor in Eq.~\ref{eq:form_factor} under a uniform external magnetic field in the $z$ direction becomes
\begin{equation*}
C=\frac{|\int_V E_z d^3x|^2}{V\int_V \epsilon(\vec{x}) |\vec{E}|^2 d^3x}.
\end{equation*}
Due to the field symmetry in the $\phi$ and $z$ directions, the integrals are simplified as 
\begin{equation}
\int_V E_z d^3x = 2 \pi L \int_0 ^R r E_z dr,
\label{eq:relation1}
\end{equation}
and
\begin{equation}
\int_V \epsilon (\vec{x}) |\vec{E}|^2 d^3x = 2 \pi L \sum_j \epsilon_j \int_j r |\vec{E}|^2 dr,
\label{eq:relation2}
\end{equation}
with $j$ denoting the three distinct regions.
The cavity quality factor is obtained from a relation
\begin{equation*}
Q = {G \over R_s},
\end{equation*}
where $G$ is the geometry factor and $R_s$ is the surface resistance.
The mode dependent geometry factor is given by
\begin{equation}
G = \mu_0 \omega {\int_V |\vec{H}|^2 d^3x \over \oint_S |\vec{H}|^2 d^2x},
\label{eq:geometry_factor}
\end{equation}
where $\mu_0$ is the vacuum permeability and $\vec{H}$ is the magnetic field strength of the resonant mode under consideration.
For normal conductors, the surface resistance in the radio frequency regime is expressed as
\begin{equation*}
R_s = {1 \over \delta \sigma},
\end{equation*}
where $\delta$ is the skin depth, $\delta\equiv\sqrt{2 / \mu_0 \sigma \omega}$, and $\sigma$ is the metal conductivity.

Using geometrical symmetries and the fact that stored electric and magnetic fields share an equal amount of energy, the volume and surface integrals in Eq.~\ref{eq:geometry_factor} become, respectively,
\begin{equation}
\int |\vec{H}|^2 d^3x = {1 \over \mu_0} \int \epsilon(\vec{x}) |\vec{E}|^2 d^3x,
\label{eq:vol_int}
\end{equation}
and 
\begin{equation}
\oint |\vec{H}|^2 d^2x = 2 \pi R L |\vec{H} (r=R)|^2 + 2 \int_{\rm end} |\vec{H}|^2 d^2x.
\label{eq:surf_int}
\end{equation}
The last term in Eq.~\ref{eq:surf_int} is for the end surfaces and is further developed to
\begin{equation*}
\int_{\rm end} |\vec{H}|^2 d^2x = {1 \over L} \int |\vec{H}|^2 d^3x = {1 \over L \mu_0} \int \epsilon(\vec{x}) |\vec{E}|^2 d^3x.
\end{equation*}
By plugging Eq.~\ref{eq:relation2} into Eqs.~\ref{eq:vol_int} and~\ref{eq:surf_int}, we obtain 
\begin{equation*}
\int |\vec{H}|^2 d^3x = {2 \pi L \over \mu_0} \sum_j \epsilon_j \int_j r |\vec{E}|^2 dr
\end{equation*}
\begin{equation*}
\oint |\vec{H}|^2 d^2x = 2 \pi R L |\vec{H} (r=R)|^2 + {4 \pi  \over \mu_0} \sum_j \epsilon_j \int_j r |\vec{E}|^2 dr.
\end{equation*}

\begin{figure}[h]
	\hspace{-0.6cm}
    \subfloat[]{
	\includegraphics[width=0.55\textwidth]{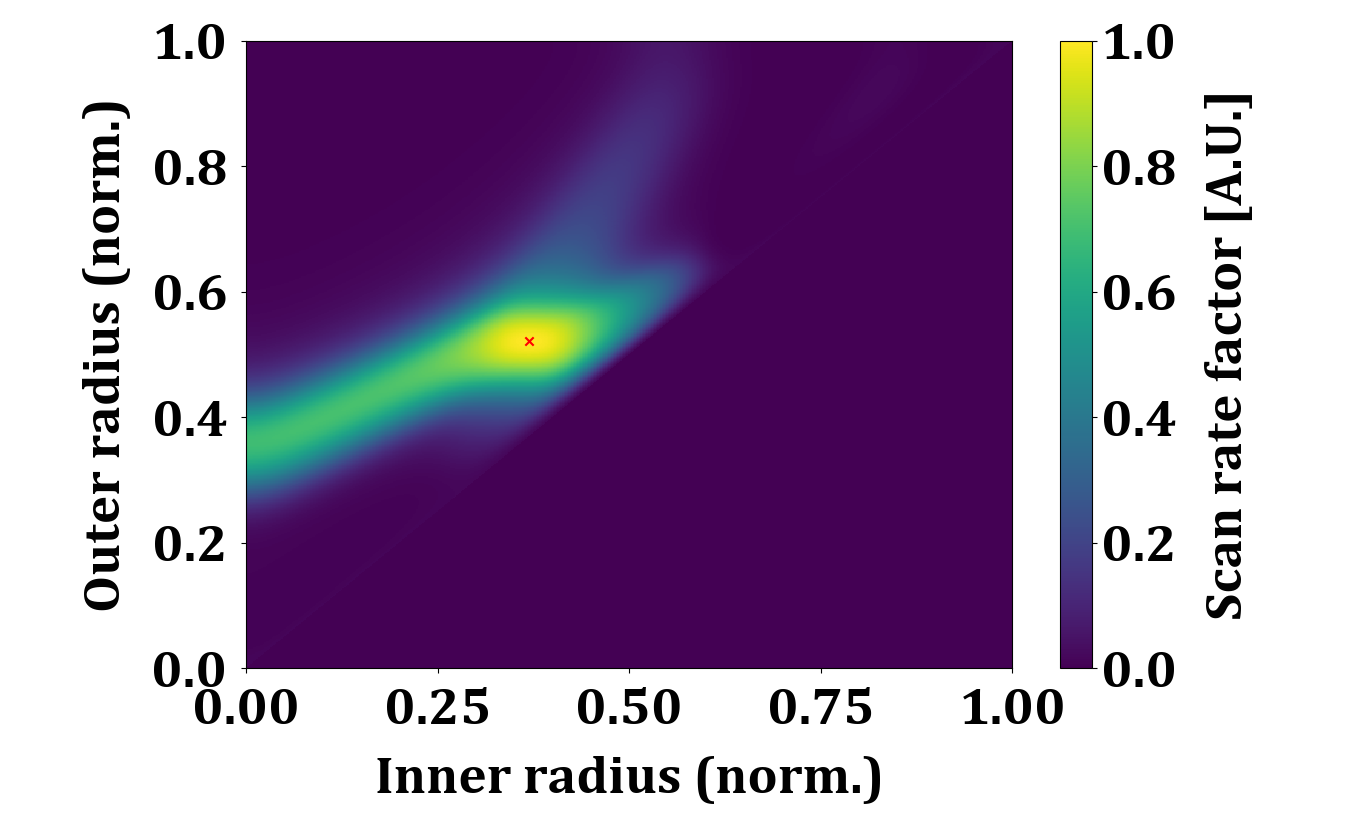}
	}
	\hspace{-0.7cm}
    \subfloat[]{	
	\includegraphics[width=0.55\textwidth]{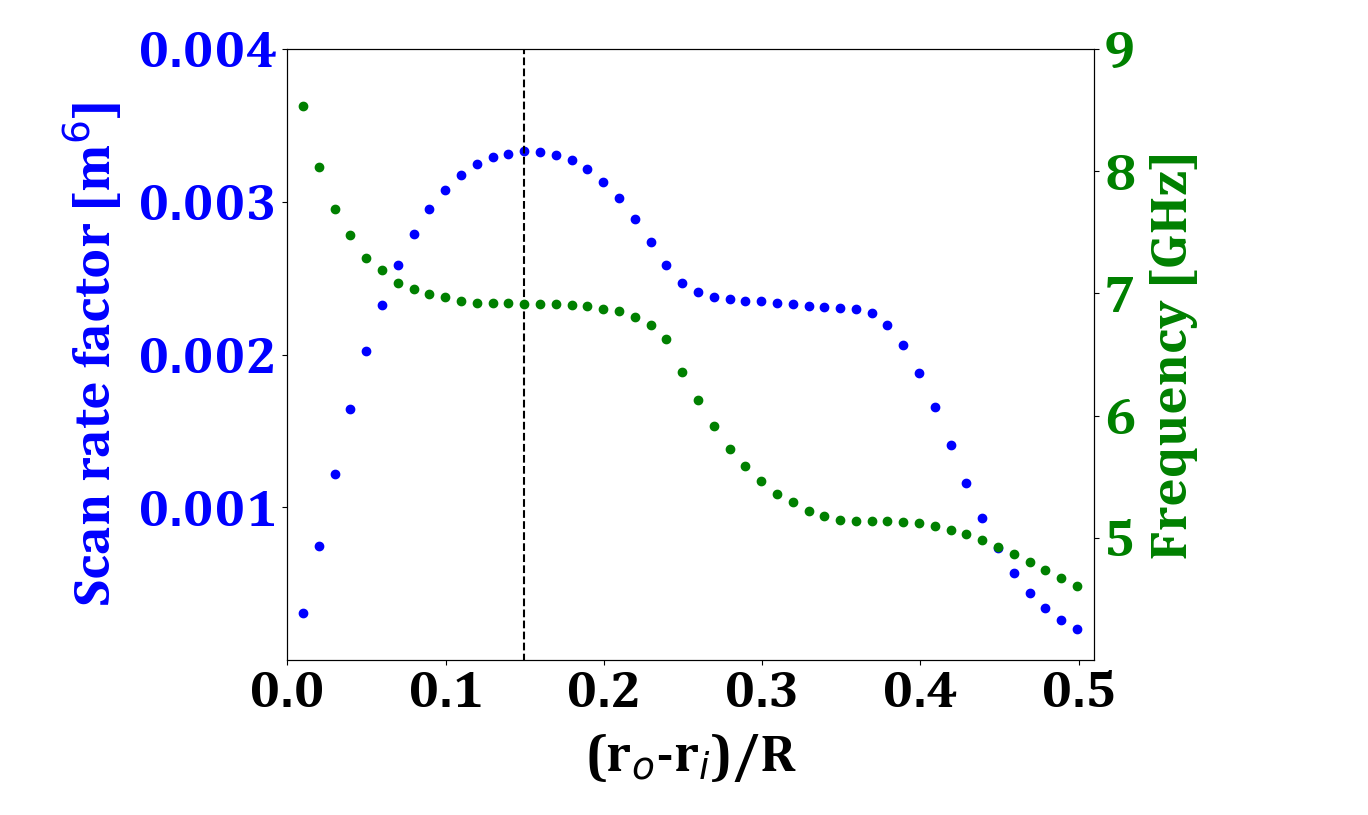}
	}
	\caption{(a) Two-dimensional plot of the scan rate factor as a function of the inner and outer radii of a dielectric cylindrical hollow.
	 Mathematica was used to compute $V^2C^2Q$ along with the EM solutions driven in the text.
	The red cross, which reads $r_i=0.37$ and $r_o=0.52$, represents the optimal point that maximizes the scan rate factor.
	(b) Scan rate factor and resonant frequency as a function of the thickness (relative to the cavity radius $R$) of the dielectric hollow with $\epsilon=10$.
	The dashed line represent the optimal thickness ($\lambda/2\sqrt{\epsilon}$), which yields the highest sensitivity.}
	\label{fig:optimization}
\end{figure}

It is noted that all the integrals required to calculate the scan rate factor, $V^2C^2Q$, are expressed in two forms:
\begin{equation*}
\int_0 ^R r E_z dr \; {\rm and} \int_j r |\vec{E}|^2 dr.
\end{equation*}
It is straightforward to compute these integrals by using the electric field solutions obtained in Sec.~\ref{sec:em_solution}.
The optimal radii of the dielectric hollow, $r_i$ and $r_o$, are found by requiring $d(V^2C^2Q)/dr=0$, which unfortunately cannot be resolved analytically.
Instead, we obtain the optimal values from a numerical approach using Wolfram Mathematica computer program~\cite{tool:mathematica}, as shown in Fig.~\ref{fig:optimization}(a).
It is found that the optimal thickness of a dielectric hollow with $\epsilon$ corresponds to approximately $\lambda/2\sqrt{\epsilon}$, where $\lambda$ is the wavelength of the EM wave of the TM$_{030}$ resonant mode with the dielectric hollow placed, which is approximately equal to the cavity radius $R$, as can be inferred from Fig.~\ref{fig:dielectric_effect}(a)\footnote{For reference, the wavelength of the TM$_{030}$ mode for an empty cylindrical cavity with a radius $R$ is given by $\lambda_{030}=2\pi R/\chi_{03}=0.726R$, where $\chi_{03}=8.654$ is the 3rd zero of the Bessel's function $J_0(kr)$.}.
We found that this optimized dimension is consistent with that determined by a simulation study, using COMSOL Multiphysics$\textsuperscript{\textregistered}$ software~\cite{tool:comsol}, where we modeled a cylindrical cavity with a 90\,mm diameter and 100\,mm height, and a dielectric hollow with $\epsilon=10$.
Figure~\ref{fig:optimization}(b) shows the dependence of the scan rate factor and resonant frequency on the thickness of the hollow.
The presence of the optimized dielectric structure makes an enhancement to the form factor by almost an order of magnitude (from 0.05 to 0.45), resulting in a substantial improvement in scan rate (about 60 times higher than that in its absence).

\section{Frequency tuning mechanisms}
\label{sec:mechanisms}
Since the axion mass (equivalently the frequency of the converted photon) is a priori unknown, a cavity must be tunable to cover a wide frequency range for a given cavity dimension.
In this section, we examine plausible frequency tuning mechanisms for the TM$_{030}$ mode utilizing the dielectric structure depicted in Sec.~\ref{sec:analytic_solution}.
We consider symmetry breaking of the electric field along the longitudinal, radial and azimuthal directions of the cylindrical coordinate system.

\subsection{Along the longitudinal direction}
\label{subsec:longitudinal}
Recently, there was a study that showed a potential application of Bragg reflectors with a dielectric hollow using the TM$_{030}$ mode of cylindrical cavities~\cite{paper:supermode}.
The study presented a frequency tuning scheme which breaks the field symmetry along the cavity axis by splitting a dielectric cylindrical hollow in the longitudinal direction.
Since the performance of the mechanism was already evaluated in that study, no attempt is made in this article for any further studies.

\subsection{Along the radial direction}
\label{subsec:radial}
\begin{figure}[h]
	\centering
		\includegraphics[width=0.2\textwidth]{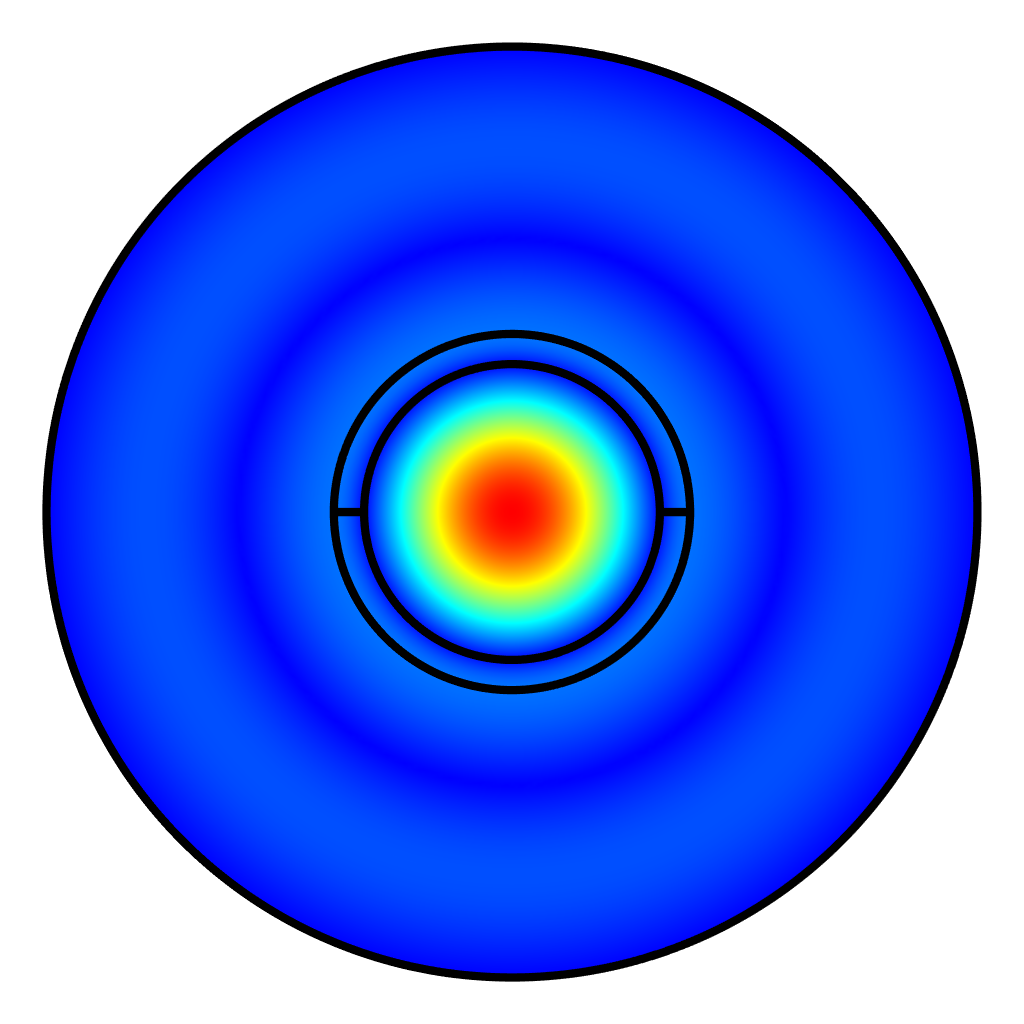}
		\includegraphics[width=0.2\textwidth]{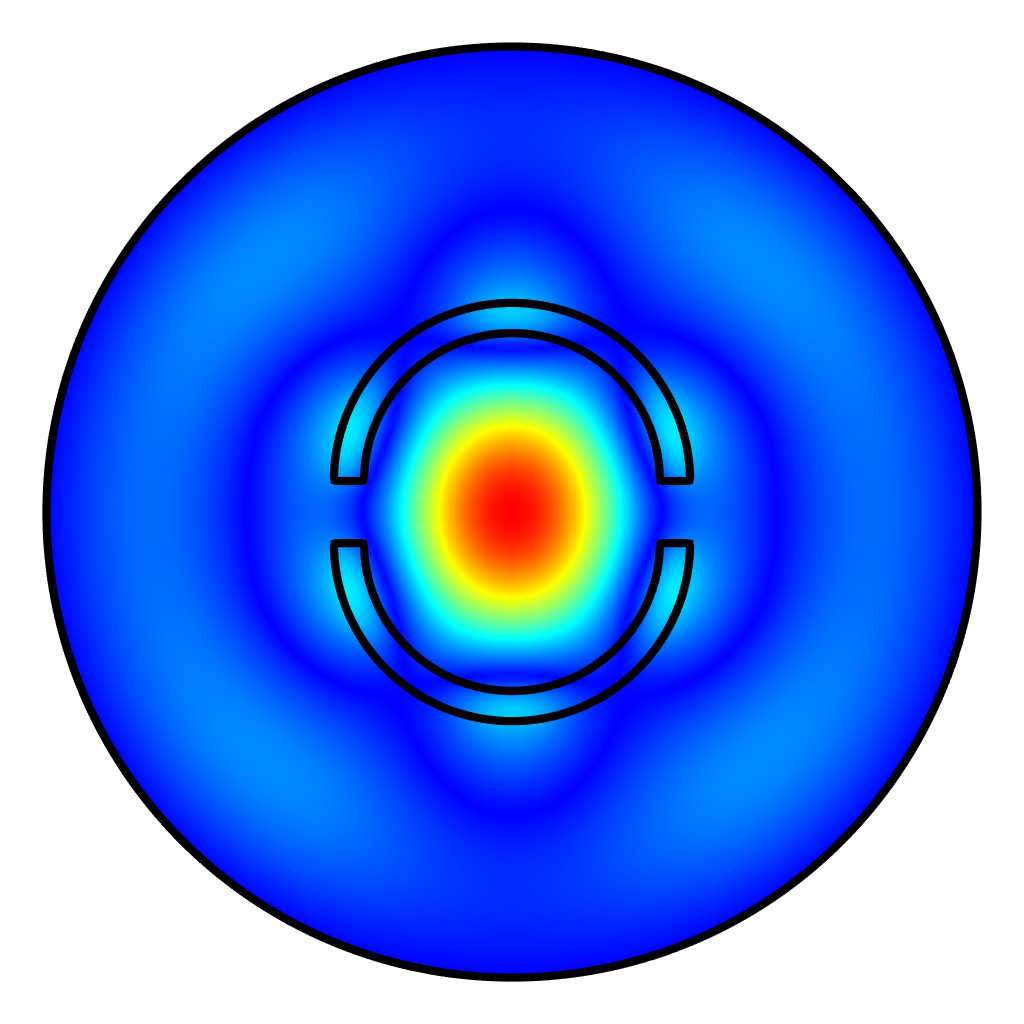}
		\includegraphics[width=0.2\textwidth]{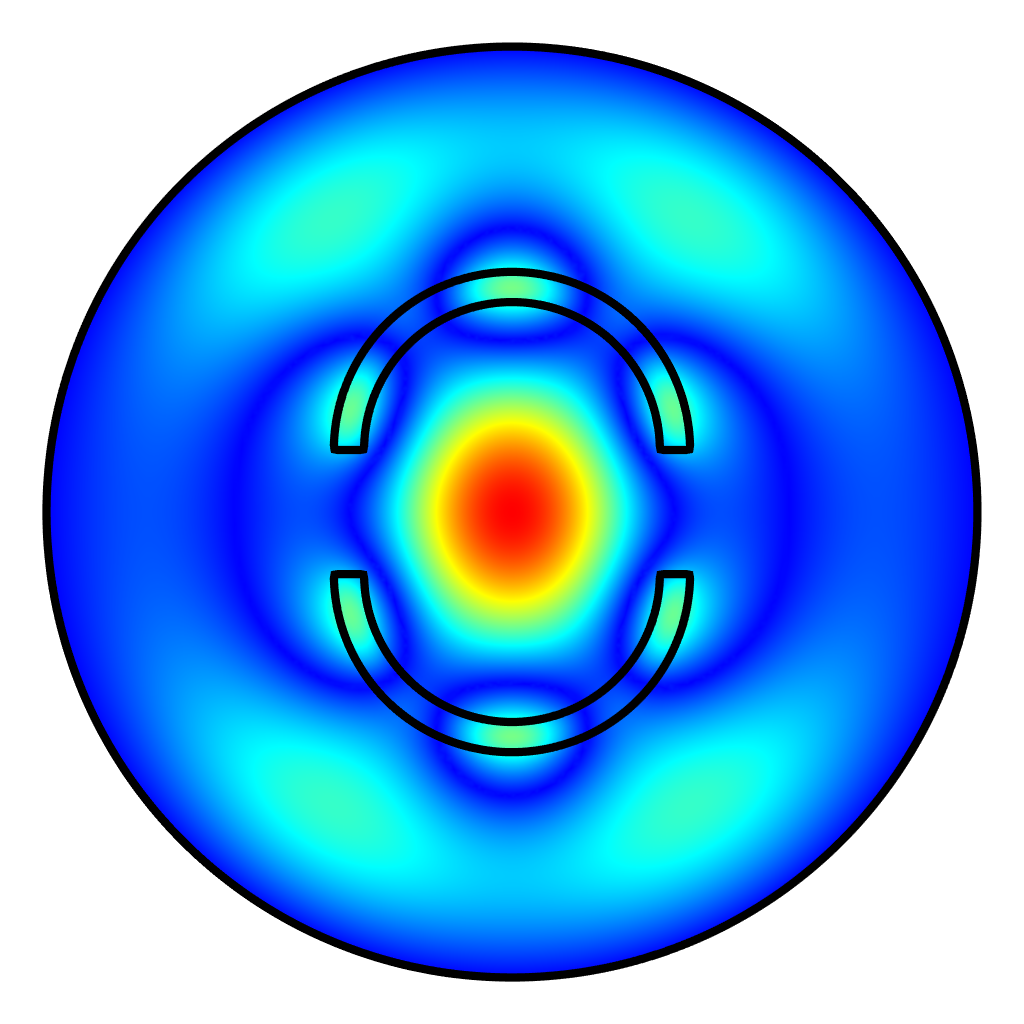}
	\caption{Illustration of the tuning scheme along the radial direction using two half-hollow cylinders.
	The color represents the strength of the electric field for the TM$_{030}$ mode.}
	\label{fig:tuning_radial}
\end{figure}

A conventional way to tune the resonant frequency of a cylindrical cavity is to break the transverse symmetry by translating a (pair of) dielectric or conducting rod(s) in the radial direction~\cite{paper:ADMX,paper:HAYSTAC}.
For a cylindrical dielectric cavity, such symmetry breaking can be achieved by splitting the dielectric hollow vertically into pieces and moving them apart along the radial direction.
For simplicity, a two-piece scheme, which is illustrated in Fig.~\ref{fig:tuning_radial}, is considered in this study.

\begin{figure}[h]
	\centering
	\includegraphics[width=0.6\textwidth]{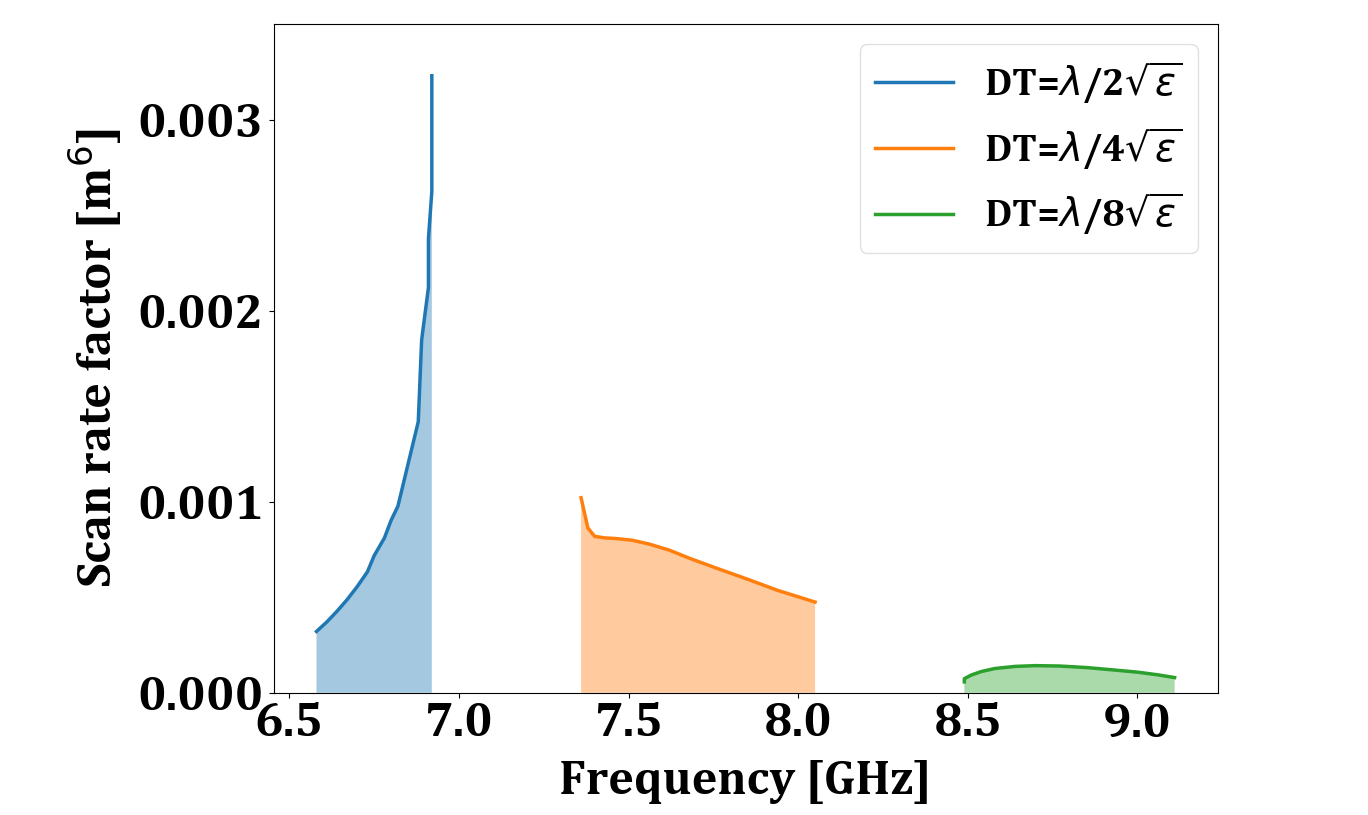}
	\caption{Simulation results of scan rate factor and tunable frequency range for different dielectric thicknesses (DTs) relying on the radial tuning mechanism.
	The same cavity dimension and dielectric property as in Sec.~\ref{subsec:scan_rate_factor} were used.
	The areas in color, corresponding to the figure of merit newly defined in the text, are 0.29 (blue), 0.47 (orange) and 0.07 (green), in unit of m$^6\cdot$MHz, respectively.
	}
	\label{fig:thickness_depend2}
\end{figure}

The initial size of the hollow cylinder is chosen from the value derived in Sec.~\ref{sec:analytic_solution}.
We found however that even though a thickness of $\lambda/2\sqrt{\epsilon}$ yielded the highest sensitivity, it resulted in a lack of tunability.
A dedicated simulation study, some results of which are shown in Fig.~\ref{fig:thickness_depend2}, indicates that the sensitivity and tuning range need to be compromised.
By defining a new figure of merit as the product\footnote{Strictly speaking, the new figure of merit is an integration of the scan rate factor over the tuning range, i.e., $\int_{\Delta f} (V^2C^2Q) df$.} of $V^2C^2Q$ and $\Delta f$, the frequency tuning range, we find that $\lambda/4\sqrt{\epsilon}$ is the optimal thickness with reasonable tunability ($\sim9\%$ with respect to the central frequency) and high sensitivity ($\sim80\%$ of that for $\lambda/2\sqrt{\epsilon}$ on average).
The maximum tuning range is achieved when the two half cylinders are separated by approximately $\lambda/2\sqrt{\epsilon}$.
However, the mechanical design for displacing the long hollow pieces in the radial direction could be challenging to implement in such sensitive experiments.

\subsection{Along the azimuthal direction}
\label{subsec:azimuthal}
\begin{figure}[h]
	\centering
		\includegraphics[width=0.2\textwidth]{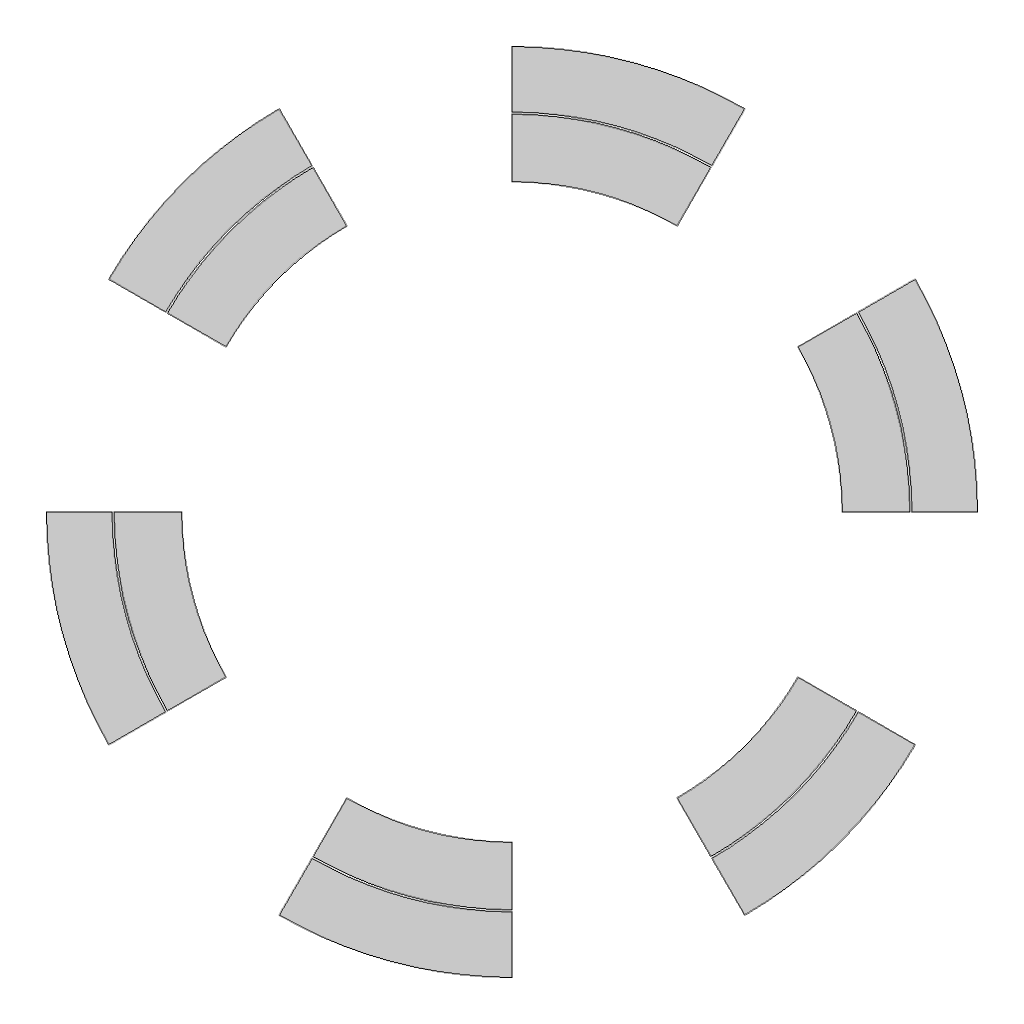}
		\includegraphics[width=0.2\textwidth]{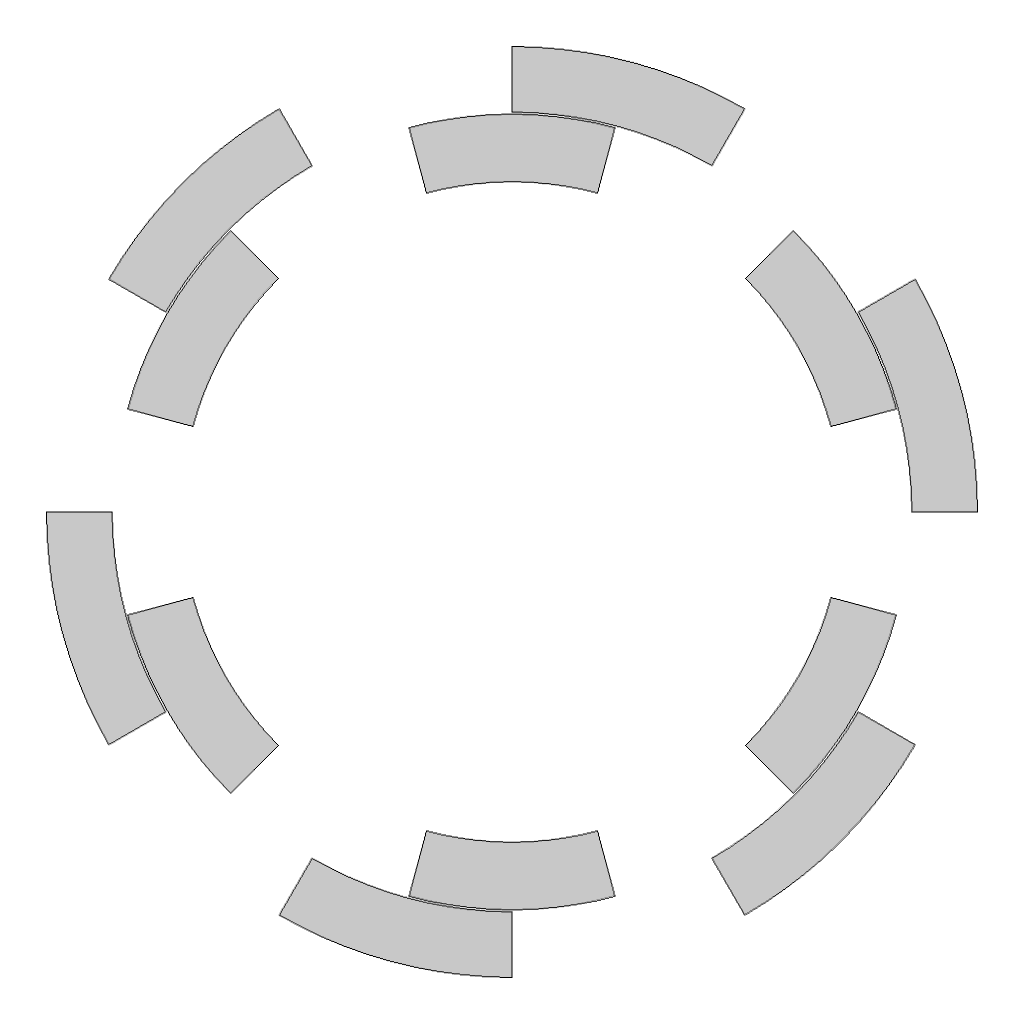}
		\includegraphics[width=0.2\textwidth]{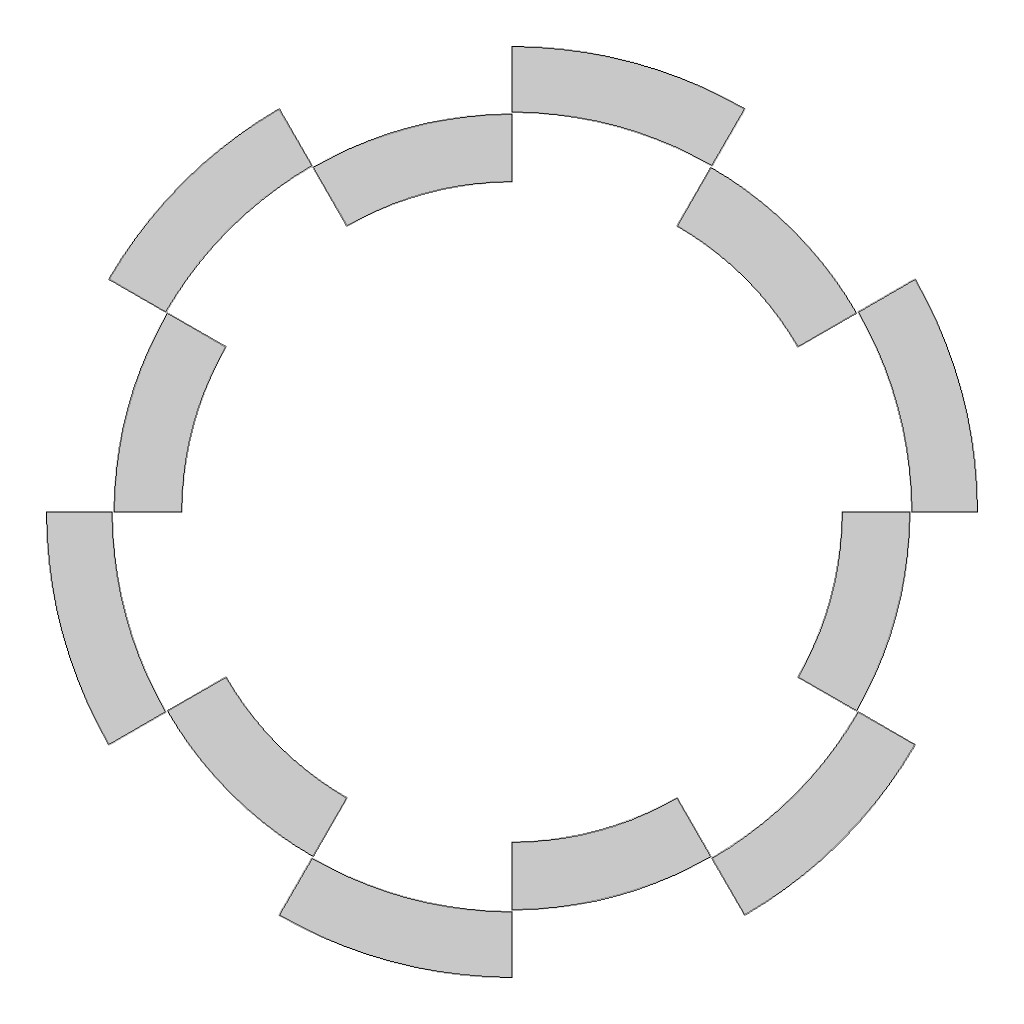}
	\caption{Illustration of the tuning mechanism described in the text. 
	The inner layer of the segments is simultaneously rotated counter clockwise with respect to the fixed outer layer.}
	\label{fig:tuning}
\end{figure}

\begin{figure}[h]
	\centering
		\includegraphics[width=0.2\textwidth]{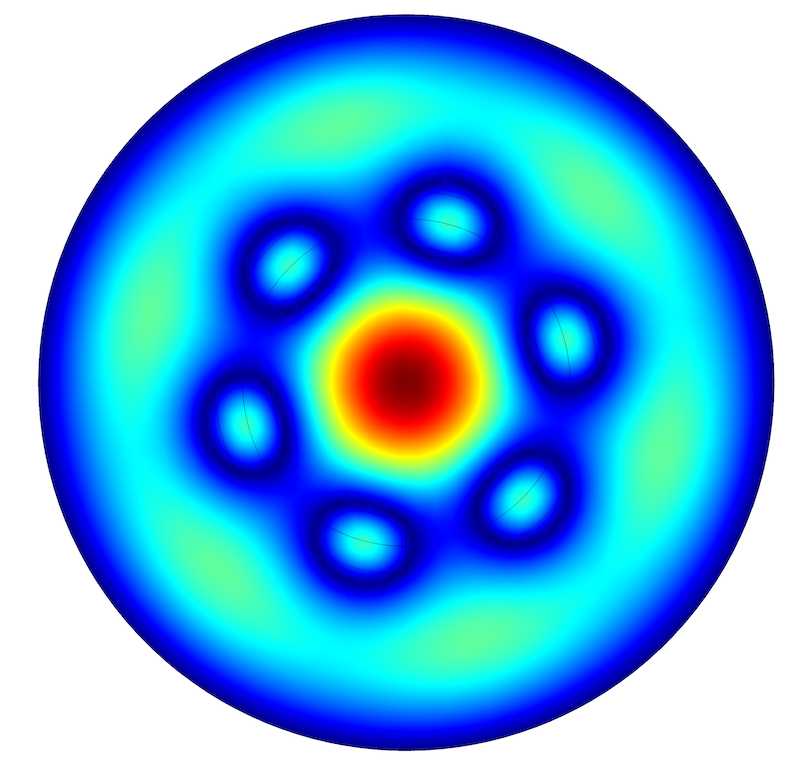}
		\includegraphics[width=0.2\textwidth]{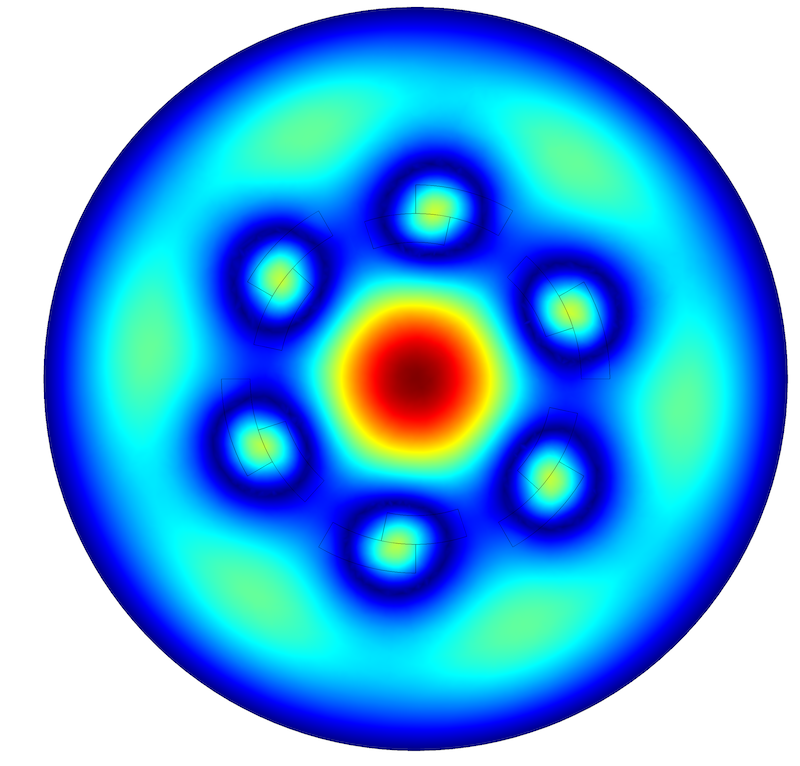}
		\includegraphics[width=0.2\textwidth]{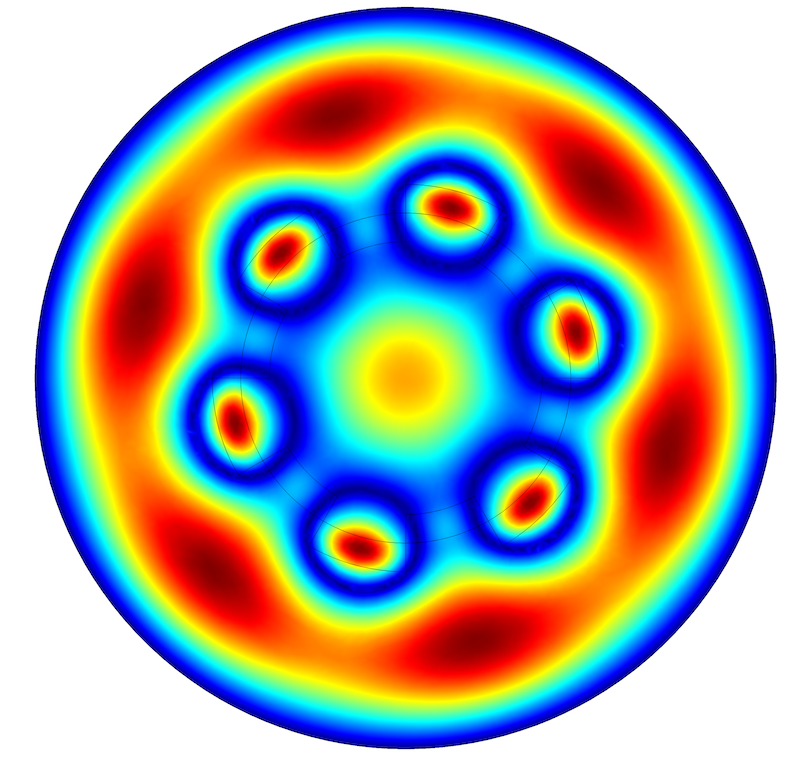}
	\caption{Electric field distributions of the TM$_{030}$-like mode with the 6-segment tuning mechanism applied to suppress the negative field component.
	Each distribution corresponds to the rotational angle of the inner layer in Fig.~\ref{fig:tuning}.}
	\label{fig:efield}
\end{figure}

The study performed in Sec.~\ref{subsec:scan_rate_factor} showed that the resonant frequency has a strong dependence on the thickness of the dielectric hollow, as seen in Fig.~\ref{fig:optimization}(b).
Based on this feature, we designed a new tuning mechanism, which consists of a double layer of concentrically segmented dielectric hollow pieces, with one layer of the segments rotating with respect to the other.
In this scheme, the outer layer of dielectric segments is fixed in position, while the inner layer is turned along the azimuthal direction simultaneously by a single rotator.
Figure~\ref{fig:tuning} visualizes how this scheme alters the effective thickness of the dielectric hollow over the course.
The corresponding electric field distributions are also shown in Fig.~\ref{fig:efield}. 
This tuning mechanism concept essentially relies on the transformation of continuous rotational symmetry into a discrete rotation symmetry, to create a tuning capability.
It is noted that since the resonant frequency is tunable using a single rotator, this design provides a more reliable tuning mechanism, one which is also easy to implement.

The design of the tuning system was optimized based on simulation studies by considering three parameters: 1) number of segments; 2) thickness of the double layer; and 3) inner radius of the inner layer.
Assuming that there were no (or weak) correlations between one another, the parameters were scanned independently to find the optimal values which maximized the aforementioned figure of merit, $\int_{\Delta f} (V^2C^2Q) df$.
Using a model of a copper cylindrical cavity with dielectric segments of $\epsilon=10$, it was found that a six segment design gives the best performance.
The optimal thickness of the layer pair and the radius of the inner layer were obtained to be approximately $\lambda/2\sqrt{\epsilon}$ and 0.37$R$, which are consistent with the analytically calculated values.
The relative thickness between the inner and outer layers was also fine tuned, such that a thicker inner layer relative to the outer layer yielded a slightly better tunability, with the overall sensitivity remaining intact.
We observed that the frequency tuning range of the TM$_{030}$ resonant mode was about 6\% with respect to the central frequency and the form factor was enhanced to greater than 0.33 (compared to 0.05 for an empty cavity) over the entire tuning range.

\subsection{Performance comparison}
\begin{figure}[h]
	\centering
	\includegraphics[width=0.65\linewidth]{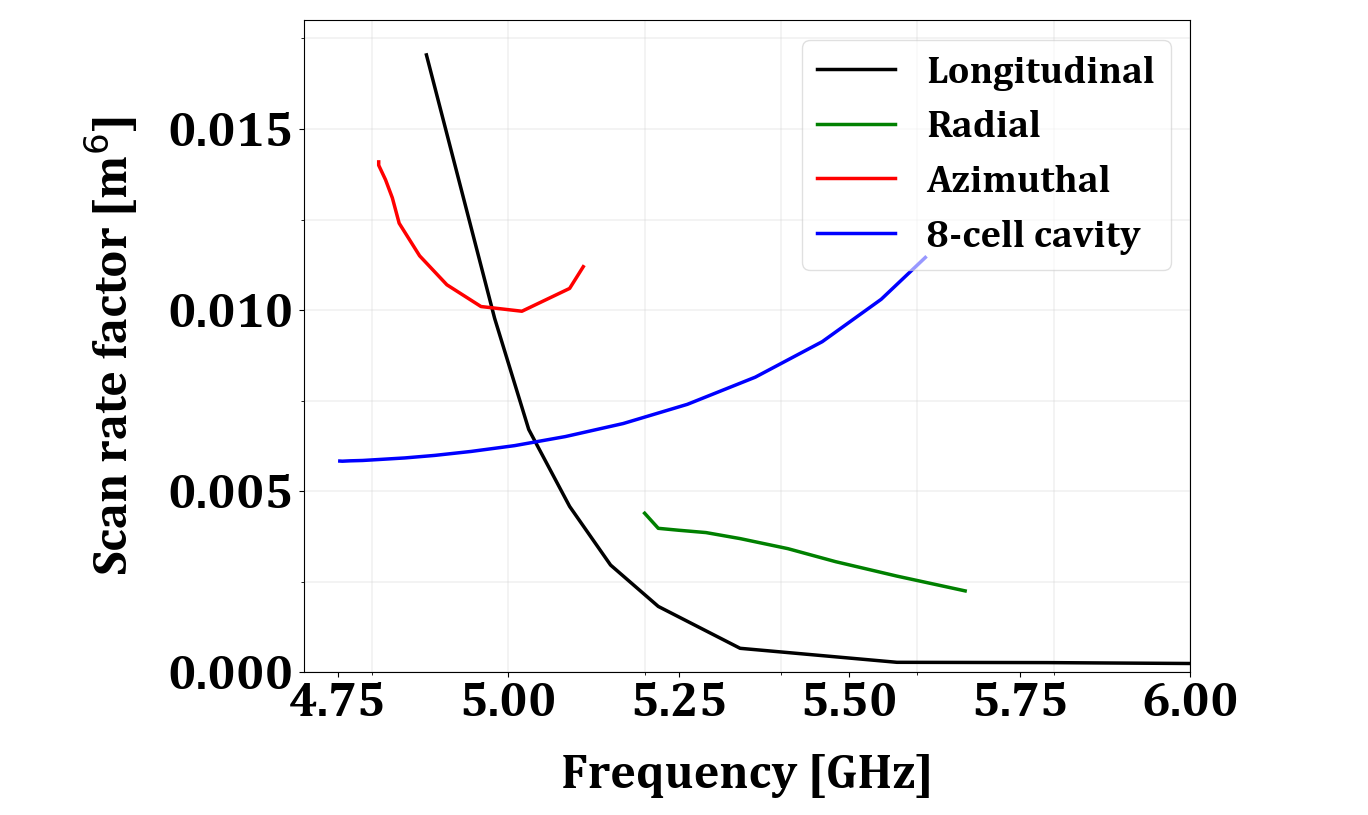}
	\caption{Performance comparison of the tuning mechanisms depicted in the text for the TM$_{030}$ mode.
	The area below each line corresponds to the figure of merit defined in the literature.
	The equivalent performance for an octuple-cell cavity~\cite{paper:multiple-cell}, which relies on the TM$_{010}$ mode, is also compared.}
	\label{fig:comparison_tm030}
\end{figure}

Sections~\ref{subsec:longitudinal}$-$\ref{subsec:azimuthal} describe the tuning schemes for the TM$_{030}$ mode, based on symmetry breaking along the longitudinal, radial, and azimuthal directions.
Using the COMSOL simulation tool, the performance of the individual mechanisms was evaluated in terms of scan rate factor and frequency tunability.
We modeled a cylindrical cavity with the same dimension as in Ref.~\cite{paper:supermode} to reproduce the results for the longitudinal mechanism to check consistency.
For each scheme, we introduced a tuning system with optimal dimensions and repeated the evaluation.
A comparison of the performance for the different mechanisms is shown in Fig.~\ref{fig:comparison_tm030}.
Among those, the mechanism along the azimuthal direction was chosen for further study, as it yields the high sensitivity over a reasonable tuning range.

\section{Experimental demonstration}
\label{sec:demonstration}
\begin{figure}[!ht]
    \centering
    \subfloat[]{
       \includegraphics[width=0.25\textwidth]{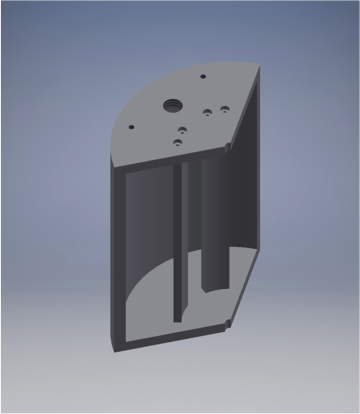}
     }
    \subfloat[]{
       \includegraphics[width=0.25\textwidth]{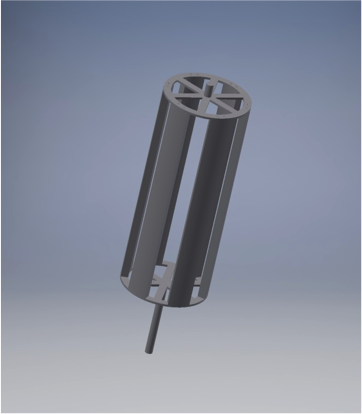}
       }
    \subfloat[]{
       \includegraphics[width=0.26\textwidth]{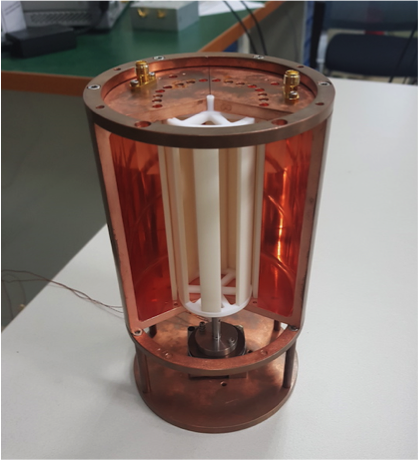}
     }
    \caption{Cavity design. 
    (a) One of three pieces of the split cavity with two outer dielectric segments fixed in place.
    (b) Assembled inner layer composed of six smaller segments and a pair of wheel-shaped structures.
    (c) Photo of the partially assembled cavity with the tuning system mounted.}
    \label{fig:design}
\end{figure}

A cavity and a tuning system were fabricated to demonstrate the experimental feasibility of the chosen tuning mechanism.
We employed a split-type cavity design, which was initially introduced in Ref.~\cite{paper:magnetoresistance}.
Made of oxygen-free high conductivity copper, the cavity consists of three identical pieces.
The assembly builds a cylindrical cavity with a 90\,mm inner diameter and 100\,mm inner height and introduces a narrow hole through the center of both the top and bottom ends.
Inside each cavity piece, two outer dielectric segments were placed at fixed positions, as shown in Fig.~\ref{fig:design} (a).
Highly pure aluminum oxide (99.7\% Al$_2$O$_3$ with $\epsilon=9.8$) was chosen as the dielectric material. 
Six smaller dielectric pieces, supported by a pair of wheel-shaped structures at the top and bottom, composed the inner layer of the tuning system, as shown in Fig.~\ref{fig:design} (b). 
The support structure was made of polytetrafluoroethylene (PTFE), which has a low dielectric constant ($\epsilon=2.1$) and a low dissipation factor (on the order of $10^{-6}$ at 4\,K).
The bottom structure was fabricated to have an extended rod piece in the middle, which was attached through the cavity hole to a single rotational piezo actuator outside the cavity to simultaneously turn the inner layer.
The overall structure of the tuning system in the partially assembled cavity is seen in Fig.~\ref{fig:design} (c).
Due to the similarity in shape and the way it works, the mechanism is dubbed a {\it wheel} mechanism.

\begin{figure}[h]
	\centering
	\includegraphics[width=0.65\linewidth]{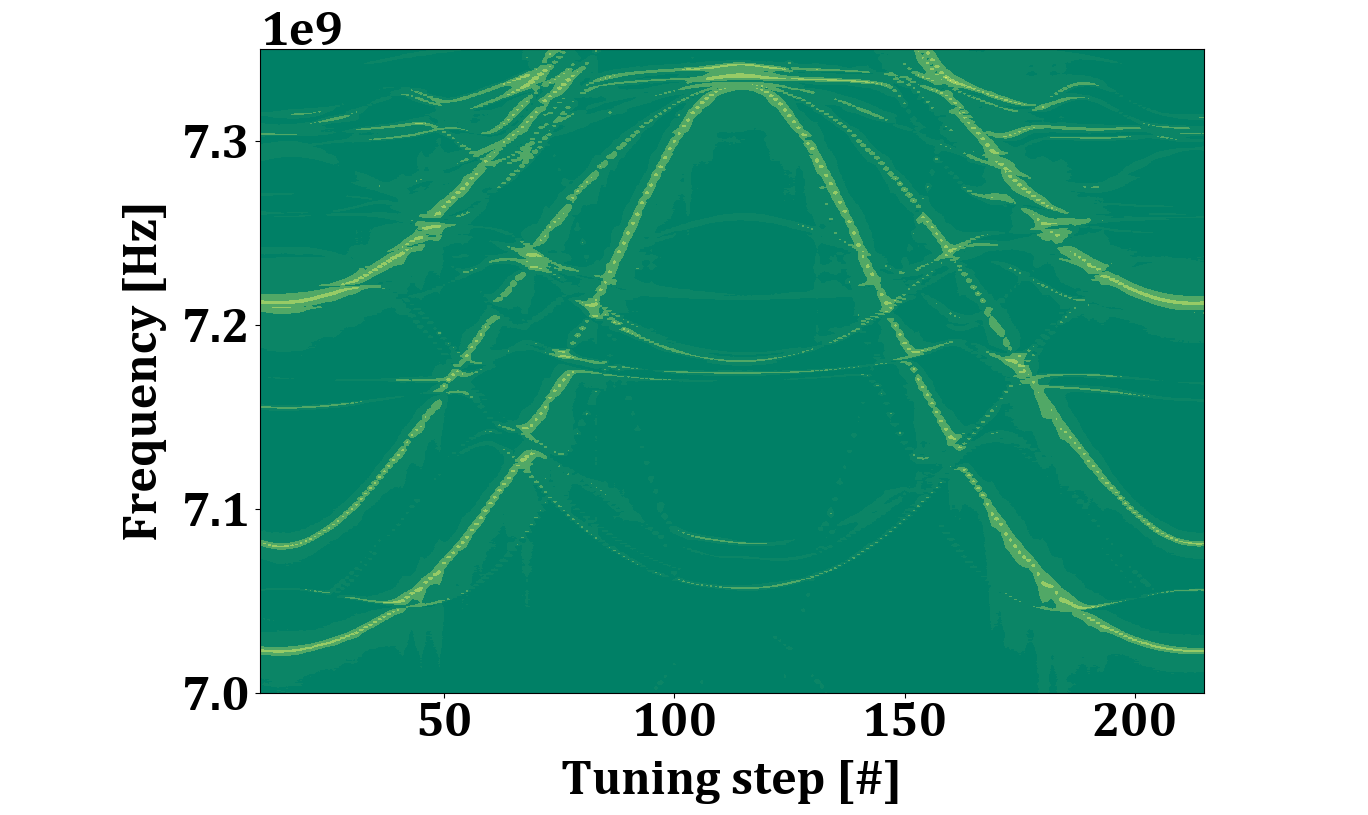}
	\caption{Measured frequency map of the cylindrical cavity over a single tuning cycle. 
	The bell-shaped curve with its tuning range between 7.02 and 7.32\,GHz corresponds to the TM$_{030}$ mode. 
	A few mode crossings by TE modes are also observed.}
	\label{fig:mode_map}
\end{figure}

The assembled cavity was installed in a cryogenic system and brought to a low temperature of around 4\,K.
The resonant frequencies and quality factors were measured using a network analyzer, through transmission signal between a pair of monopole RF antennae weakly coupled to the cavity.
A frequency map of the resonant modes of the cavity was drawn as a function of the tuning step, while rotating the inner layer of the tuning system.
We observed the periodic behavior of the resonant modes over a full tuning process.
Figure~\ref{fig:mode_map} shows the frequency map over a single cycle, which corresponds to the rotational angle of 60$^\circ$.
The bell-shaped curve with its frequency spanning from 7.02 to 7.32\,GHz corresponds to the TM$_{030}$ mode.
This frequency region is about three times higher than the TM$_{010}$ resonant frequency of the same cavity, $f_{010}=2.55$\,GHz.
Symmetric and smooth frequency curves over a tuning range of $\sim$300\,MHz indicates the stability of the tuning mechanism with reasonable tunability.
The measured quality factors varied between 110,000 in the low frequency regions and 90,000 for the high frequency regions, which were consistent with the simulation results. 
This verifies that the design is a plausible approach for utilizing the higher-order resonant modes for axion haloscopes.

\section{Conclusions}
We exploited higher-order resonant modes for axion haloscope experiments to extend the search range towards high mass regions.
In particular, we examined various tuning mechanisms for the TM$_{030}$ mode by introducing a structure of dielectric material, which substantially enhanced the form factor.
The general EM solutions for a cylindrical cavity were obtained analytically and found to be consistent with the numerical calculations.
Depending on the way that the field symmetry is broken, three schemes can be considered: frequency tuning along the longitudinal, radial, and azimuthal directions.
For each scheme, a tuning system was optimally designed based on simulation studies by maximizing a new figure of merit, $\int_{\Delta f} (V^2C^2Q) df$.
Among those, the tuning scheme that relies on the symmetry breaking along the azimuthal direction, showed the highest sensitivity over a reasonable tuning range with a realistic design.
The experimental feasibility of this mechanism was demonstrated using a three-piece copper cavity and a double-layer of alumina hollow segments.
We conclude that higher-order resonant modes utilizing suitable frequency tuning mechanisms are certainly applicable to haloscope searches for high mass dark matter axions.

\ack{This work was supported by IBS-R017-D1-2019-a00 / IBS-R017-Y1-2019-a00.}


\end{document}